\documentclass[
 reprint,
 superscriptaddress,
 amsmath,amssymb,
 aps,
 floatfix,
]{revtex4-2}

\usepackage{color}
\usepackage{graphicx}
\usepackage{dcolumn}
\usepackage{bm}

\begin{document}


\title{Statistical properties of COVID-19 transmission intervals in Republic of Korea}

\author{Yeonji Seo}
\affiliation{Department of Physics, The Catholic University of Korea, Bucheon 14662, Republic of Korea}

\author{Okyu Kwon}
\affiliation{National Institute for Mathematical Sciences, Daejeon 34047, Republic of Korea}

\author{Hang-Hyun Jo}
\email{h2jo@catholic.ac.kr}
\affiliation{Department of Physics, The Catholic University of Korea, Bucheon 14662, Republic of Korea}

\date{\today}

\begin{abstract}
A transmission interval for an infectious disease is important to understand epidemic processes in complex networks. The transmission interval is defined as a time interval between one person's infection and their infection to another person. To study statistical properties of transmission intervals, we analyze a COVID-19 dataset of confirmed cases in Republic of Korea that has been collected for two years since the confirmation of the first case on 19 January 2020. Utilizing demographic information of confirmed individuals, such as sex, age, residence location, and the nature of relation between infectors and infectees, we find that transmission intervals are rarely affected by sexes, but they tend to have larger values for the youngest and oldest age groups than other groups. We also find some metropolitan cities or provinces with relatively larger (smaller) transmission intervals than other locations. These empirical findings might help us to better understand dynamical mechanisms of epidemic processes in complex social systems.
\end{abstract}

\maketitle

\section{Introduction}

In recent years epidemic processes in complex networks have been extensively studied in various fields including physics~\cite{Pastor-Satorras2015Epidemic}. The network structure has been known to have significant impact on the dynamics of epidemic processes, such as disease spreading, that take place in such networks. More recently, researchers have focused on temporal interaction patterns between individual nodes in a network that also influence the spreading behavior~\cite{Holme2012Temporal, Karsai2011Small, Rocha2011Simulated, Jo2014Analytically, Karsai2018Bursty, Hiraoka2018Correlated, Jo2019Bursty}. One of the most basic quantities for analyzing temporal spreading patterns is a transmission interval, which is defined as a time interval from an infection of one node to the infection by the node to its neighboring node~\cite{Fine2003Interval, Zhao2020Estimating, Sender2022Unmitigated}. Thus, statistical properties of transmission intervals play an important role in understanding the mechanisms behind spreading dynamics in complex systems.

In order to investigate statistical properties of transmission intervals, we analyze the COVID-19 dataset of confirmed cases in Republic of Korea (Korea hereafter) that has been collected for two years since the first confirmation on 19 January 2020~\cite{Kwon2023Clustering}. The dataset contains temporal information, such as dates of report of confirmed individuals, as well as information on precedent confirmed individuals or infectors. Defining the transmission interval as the time interval between dates of report of infectors and infectees, one can study statistical properties of transmission intervals. In addition, demographic information for confirmed individuals, such as sex, age, residence location, and the nature of relations between infectors and infectees, is also available to a large extent. It enables us to investigate the impact of demographic information of confirmed individuals on their transmission intervals.

As a result, we find that the transmission intervals are rarely affected by sexes, but they tend to have larger values for the youngest and oldest age groups than other groups. We also find that it takes longer (shorter) time for infection to occur in Busan, Daegu, and Chungbuk (Gangwon and Jeonnam) than other locations. The distribution of transmission intervals between family members (friends) is broader (narrower) than that between coworkers, which is the most similar to the distribution of the entire transmission intervals. Finally, we study the transmission interval defined at the location level, corresponding to the duration for which COVID-19 stays or circulates within the location. These empirical findings might help us to better understand the dynamical processes of disease spreading in complex social systems.

\section{Data}

Since the first confirmed case of COVID-19 in Korea, the dataset of confirmed cases of COVID-19 has been collected by Korea Disease Control and Prevention Agency~\cite{Ko2021COVID19, Jeon2022Evolution, Shim2022Transmission}. This dataset contains 670,483 confirmed cases reported to the Korean government from 19 January 2020 to 11 January 2022. Each infected individual is represented with an ID and this ID is associated with information on sex, age, residence location at the level of basic administration regions, and date of report as well as the infector's ID and the relation with the infector. However, not all IDs are associated with such information; the residence location is not available for 72,143 IDs and the date of report is not available for 3,878 IDs. Also, the information on the infectors is available for 183,277 confirmed cases; for 120,138 confirmed cases among them, the nature of relations between infectees and infectors is available, which can be summarized to be either family, coworker, or friend. We manually corrected 26 IDs for infectors and 129 dates of report having notation errors. There are 488 cases with multiple infectors to the same infectee, in which cases one of infectors has been randomly chosen. Finally, we only consider the transmission interval between 0 and 21 days, leaving us with 179,685 transmission intervals in total.

\section{Results}

\begin{figure}[t]
  \centering
  \includegraphics[width=\columnwidth]{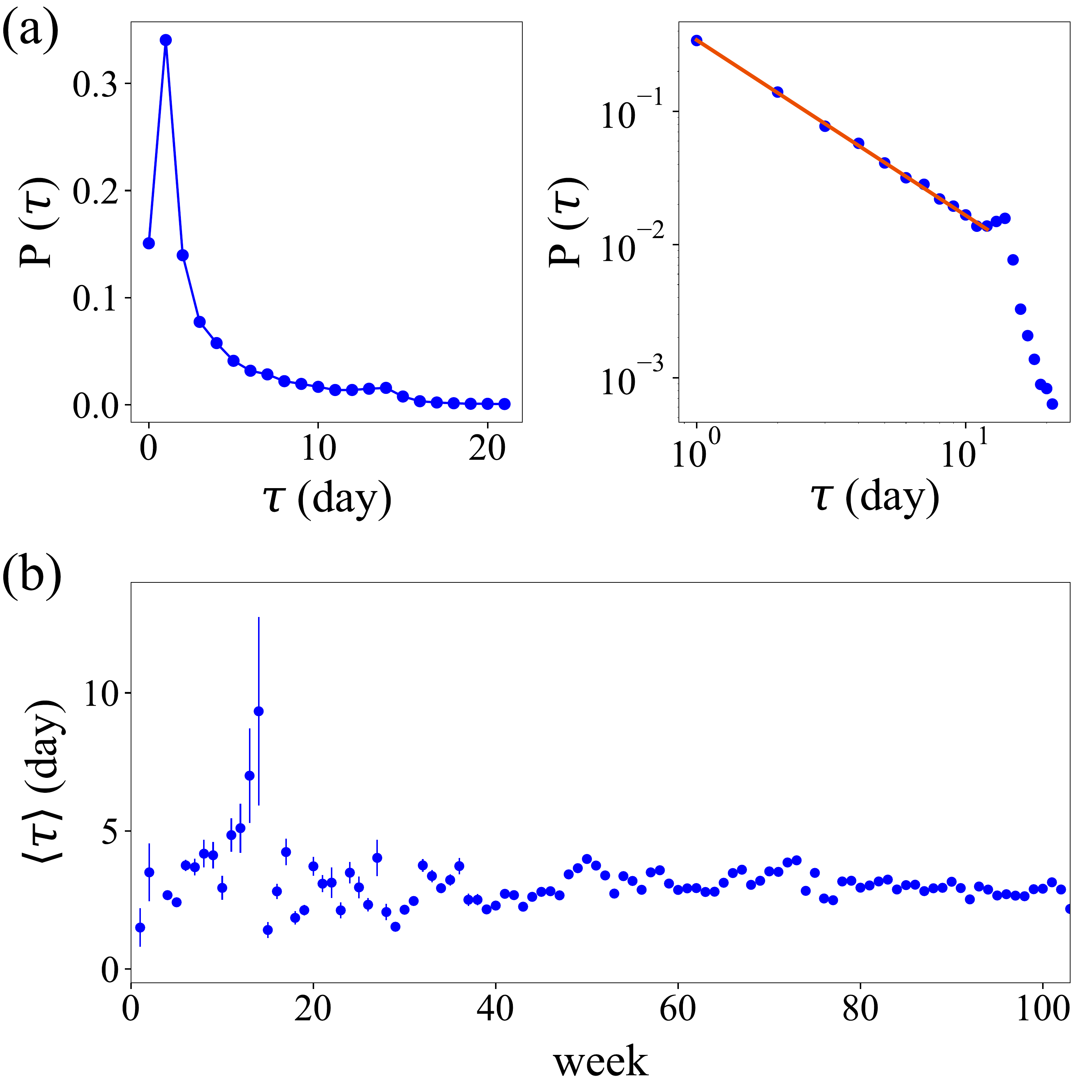}
  \caption{(a) Transmission interval distribution $P(\tau)$ for the entire dataset in a linear plot (left) and in a log-log plot (right), where the transmission interval $\tau$ is measured in days. The orange line in the right panel shows the power-law fit to $P(\tau)$ for $1\leq \tau\leq 14$.
  (b) Time series of the average transmission interval $\langle \tau\rangle$ in days, calculated for each week since 19 January 2020 [Eq.~\eqref{eq:tau_week}]. Error bars denote standard errors.
  }
  \label{fig:pdf}
\end{figure}

\subsection{Definitions}

We define terms for our paper. Each confirmed individual $i$ is associated with their sex being either male (M) or female (F), which is denoted by $s_i\in\{\rm M, F\}$. The age group the individual $i$ belongs to is set to be among 10 groups, i.e., $a_i\in \{\textrm{0--9}, \textrm{10--19}, \ldots, \textrm{80--89}, 90+\}$. Here the group of ``90+'' includes all people older than 90 years old. As for the residence locations, we mainly consider the wide-area administrative regions, namely, eight metropolitan cities, such as Seoul and Busan, and nine provinces, such as Gyeonggi and Gangwon. The residence location of the individual $i$ is denoted by $l_i\in\{\textrm{Seoul},\textrm{Busan},\ldots,\textrm{Jeju}\}$. Finally, for confirmed cases whose nature of relations between them is available, the relation between the infector $i$ and the infectee $j$ is denoted by $r_{i\to j}\in\{\textrm{Family}, \textrm{Coworker}, \textrm{Friend}\}$. Note that the relation is reciprocal, while the arrow from $i$ to $j$ only indicates the direction of infection.

\begin{figure}[t]
  \centering
  \includegraphics[width=\columnwidth]{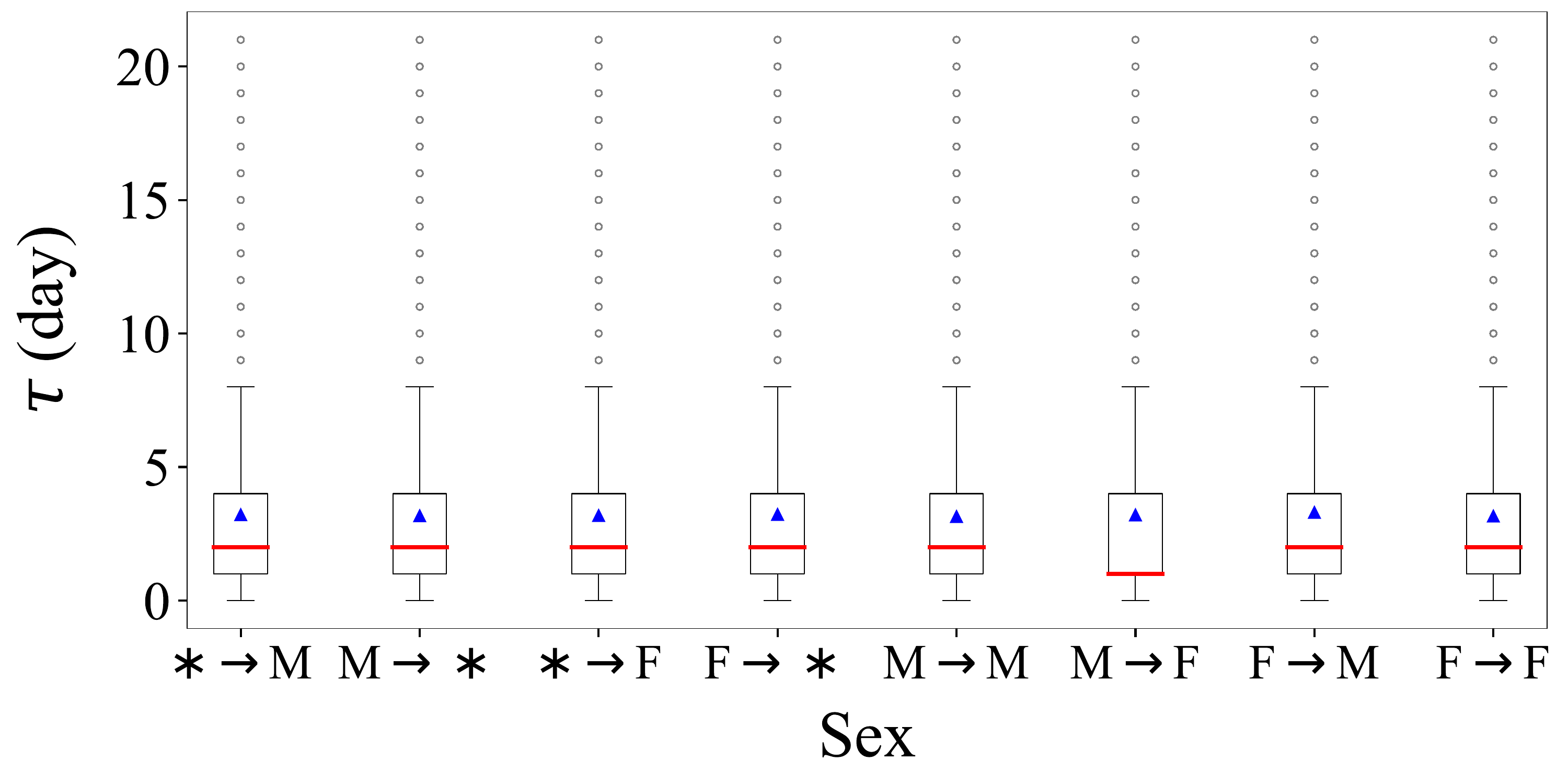}
  \caption{Box-and-whisker plots of transmission intervals $\tau$ for cases involving male/female infectors/infectees as well as for cases from male to male, from male to female, from female to male, and from female to female [Eqs.~\eqref{eq:tau_sex1}--\eqref{eq:tau_sex3}]. In every box-and-whisker plot, empty circles, blue triangle, and red line denote outliers, average, and median of $\tau$s, respectively.
  }
  \label{fig:sex}
\end{figure}

\begin{figure*}[t]
  \centering
  \includegraphics[width=\linewidth]{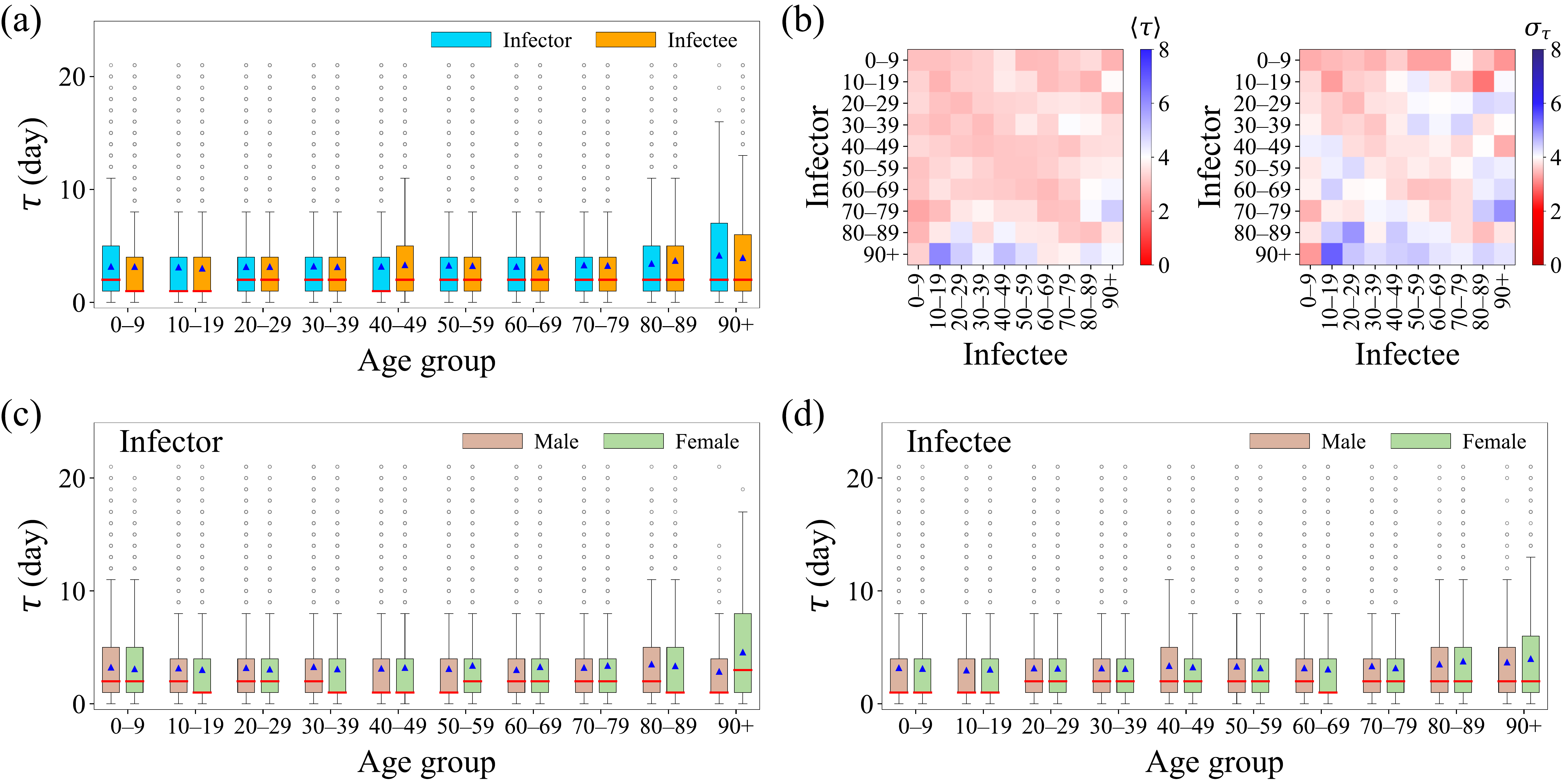}
  \caption{
  (a) Box-and-whisker plots of $\tau$s for different age groups of infectors (cyan) and infectees (orange) [Eqs.~\eqref{eq:tau_age1}~and~\eqref{eq:tau_age2}].
  (b) Heatmaps of the average (left) and the standard deviation (right) of $\tau$s between every pair of age groups of infectors and infectees [Eq.~\eqref{eq:tau_age3}].
  (c,~d) Box-and-whisker plots of $\tau$s for different age groups and sexes of infectors (c) and infectees (d) [Eqs.~\eqref{eq:tau_sexage1}~and~\eqref{eq:tau_sexage2}]; males denoted in brown and females in green. In every box-and-whisker plot, empty circles, blue triangle, and red line denote outliers, average, and median of $\tau$s, respectively.
  }
  \label{fig:age}
\end{figure*}

Each confirmed individual, denoted by $i$, with information on the infector, denoted by $j$, enables us to define an infection event $e_{j\to i}$ at time $t_i$. Here $t_i$ is the number of days elapsed since 19 January 2020 until the date of report for $i$. $t_{i}$ may not be necessarily the same as the date of infection, while it is considered as a proxy. Then a directed transmission network can be derived with 226,147 individual nodes and 179,685 directed links in total. Due to the incompleteness of the dataset, the network is not connected but made of a number of components. We leave the study on network structure for future work.

\subsection{Distribution and trend of transmission intervals}

The transmission interval is defined as a time interval between an infection of one person and their infection to another person. Precisely, if one infected individual $i$ infects another individual $j$ at time $t_j$, the transmission interval between them, denoted by $\tau_{i\to j}$, is obtained as
\begin{align}
    \tau_{i\to j}\equiv t_j - t_i.
\end{align}
From the dataset we get 179,685 transmission intervals in total, and their average and standard deviation are $\langle \tau\rangle\approx 3.2$ days and $\sigma_\tau\approx 3.8$ days, respectively.

We obtain the probability distribution function $P(\tau)$ as shown in Fig.~\ref{fig:pdf}(a). We find that the most common transmission interval is one day. Considering the fact that the COVID-19 test result typically comes out on the next day of the test in Korea, the cases with $\tau=0$ may possibly mean that infectors and infectees took the test on the same day. When plotted in a log-log scale, $P(\tau)$ seems to have a power-law regime for $\tau$ less than two weeks, which can be fitted with the estimated power-law exponent $\approx$$1.32(2)$. Then the distribution shows a small peak around at $\tau=14$ days, i.e., two weeks. It might be understood by the fact that the incubation period of COVID-19 was recognized as 14 days so that people who had close contact with confirmed individuals or migrated from abroad were supposed to stay in isolation at least for 14 days~\cite{2020COVID19}.

To see the trend of transmission intervals over time, we collect the transmission intervals for the $n$th week as following:
\begin{align}
    \mathcal{T}_n=\{\tau_{i\to j}| 7n\leq t_j<7(n+1)\}. \label{eq:tau_week}
\end{align}
Taking the average of transmission intervals for each week, we plot its trend in Fig.~\ref{fig:pdf}(b). It shows the overall decreasing tendency of transmission intervals.

\subsection{Effects of sex and age on transmission intervals}

We investigate how demographic information, i.e., sex and age, might affect the transmission interval statistics. Let us first look at the effect of sex on the transmission interval statistics. For this, we collect sets of transmission intervals involving male/female infectors/infectees, respectively, as follows:
\begin{align}
    \mathcal{T}^{(s)}_{\alpha\to *} &=\{\tau_{i\to j}| s_i=\alpha\},
    \label{eq:tau_sex1}\\
    \mathcal{T}^{(s)}_{*\to\alpha} &=\{\tau_{i\to j}| s_j=\alpha\},
    \label{eq:tau_sex2}
\end{align}
where $\alpha\in\{\textrm{M},\textrm{F}\}$, and ``$*$'' means all cases. Similarly, we get four sets of transmission intervals for all combinations of sexes of infectors and infectees as
\begin{align}
    \mathcal{T}^{(s)}_{\alpha\to \beta} =\{\tau_{i\to j}| s_i=\alpha\ \textrm{and}\ s_j=\beta\},\label{eq:tau_sex3}
\end{align}
where $(\alpha,\beta)=$ (M,M), (M,F), (F,M), and (F,F). For all eight sets of transmission intervals, we generate the box-and-whisker plots to compare the transmission interval distributions. As shown in  Fig.~\ref{fig:sex}, any significant differences between the transmission interval distributions of eight sets are not observed.

\begin{figure*}[t]
  \centering
  \includegraphics[width=\linewidth]{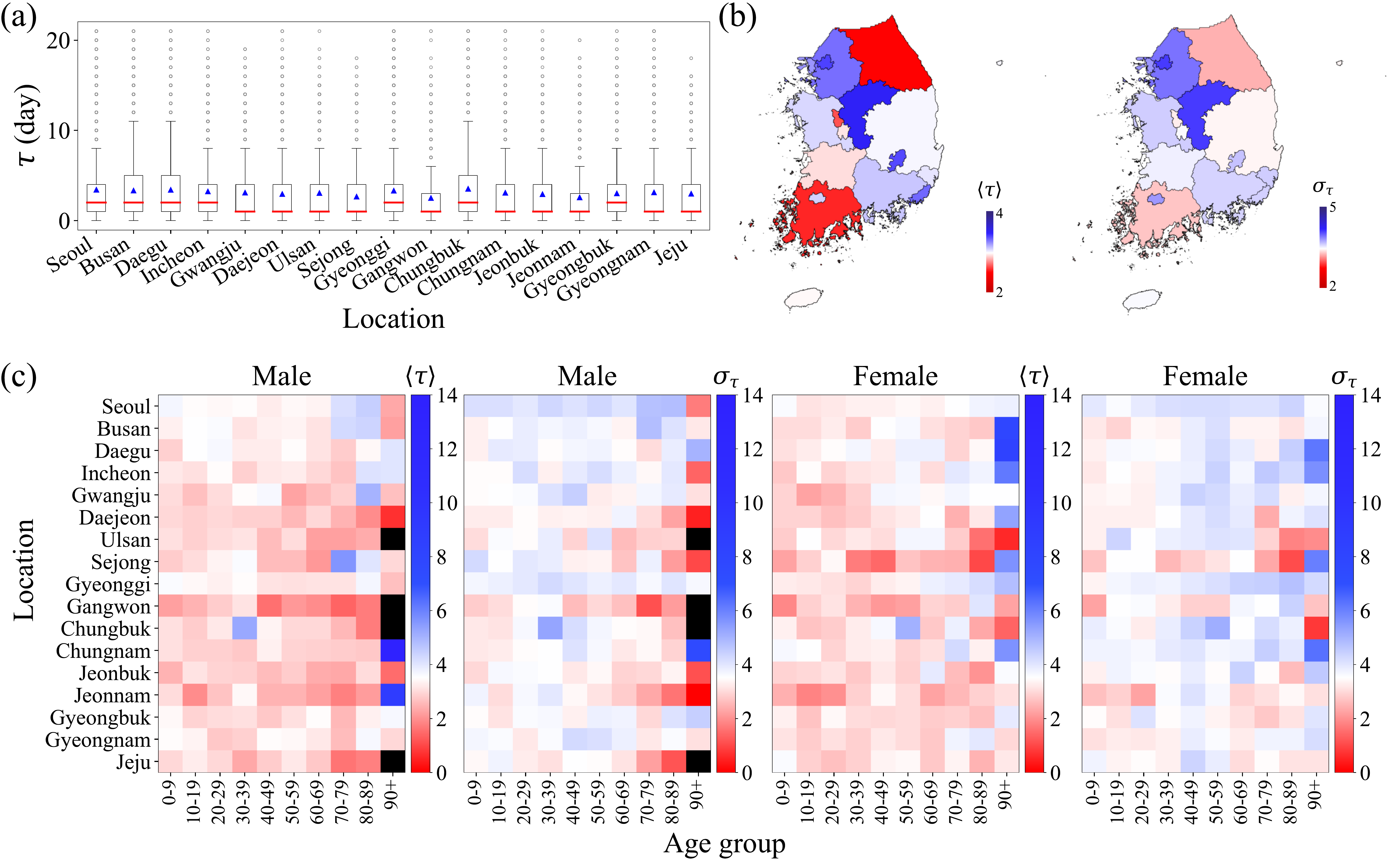}
  \caption{(a) Box-and-whisker plots of $\tau$s for different locations of infectors [Eq.~\eqref{eq:tau_loc}], where empty circles, blue triangle, and red line denote outliers, average, and median of $\tau$s, respectively.
  (b) Geographical visualization of the average (left) and the standard deviation (right) of $\tau$s for different locations of infectors. The boundary shape file for the map was downloaded from Ref.~\cite{Koreab}.
  (c) Heatmaps of the average and the standard deviation of $\tau$s for male infectors (left two panels) and female infectors (right two panels) of different locations and age groups [Eq.~\eqref{eq:tau_sexageloc}]. Black cells indicate the absence of available data.
  }
  \label{fig:loc}
\end{figure*}

Next, we study the effect of age groups on the transmission interval statistics. We define sets of transmission intervals for each age group as well as for each pair of age groups:
\begin{align}
    \mathcal{T}^{(a)}_{\alpha\to *} &=\{\tau_{i\to j}| a_i=\alpha\}, \label{eq:tau_age1}\\
    \mathcal{T}^{(a)}_{*\to\alpha} &=\{\tau_{i\to j}| a_j=\alpha\}, \label{eq:tau_age2}\\
    \mathcal{T}^{(a)}_{\alpha\to \beta} &=\{\tau_{i\to j}| a_i=\alpha\ \textrm{and}\ a_j=\beta\}, \label{eq:tau_age3}
\end{align}
where $\alpha,\beta\in\{\textrm{0--9},\ldots,\textrm{90+}\}$. Results for $\mathcal{T}^{(a)}_{\alpha\to *}$ and $\mathcal{T}^{(a)}_{*\to \alpha}$ are shown in Fig.~\ref{fig:age}(a). We find that the transmission intervals for the oldest age groups ``80--90'' and ``90+'' are overall larger than those of other groups for both infectors and infectees, while the cases with infectors of age group ``0--9'' and infectees of age group ``40--49'' show broader distributions of transmission intervals. One can at least argue that the youngest infectors and oldest infectors/infectees may not go to the test sites by themselves so that it takes more time to take the test than other age groups. 

We then calculate the average and the standard deviation of transmission intervals in each $\mathcal{T}^{(a)}_{\alpha\to \beta}$, denoted by $\langle\tau\rangle$ and $\sigma_\tau$, respectively. The results are shown as heatmaps in Fig.~\ref{fig:age}(b). Note that these calculations include transmission intervals that might be classified as outliers in the box-and-whisker plots. Also, the average and the standard deviation tend to be positively correlated with each other because transmission interval distributions are highly right-skewed. It is found that the average transmission intervals have overall similar values except for the cases involving the oldest age group ``90+''. On the other hand, the standard deviations tend to be smaller between the similar age groups and between the youngest age group and all other age groups than other cases. These findings can be interpreted such that it takes short time for infections to occur between people of similar ages or to/from kids or teenagers. It is probably because they make contact with each other more frequently or regularly, or they inform their contacts of the confirmation more quickly than expected, leading to faster testing.

Finally, we study the statistical properties of transmission intervals when considering the sex and age simultaneously. For this, we define sets of transmission intervals as follows:
\begin{align}
    \mathcal{T}^{(sa)}_{\alpha\beta\to *} &=\{\tau_{i\to j}| s_i=\alpha\ \textrm{and}\ a_i=\beta\},
    \label{eq:tau_sexage1}\\
    \mathcal{T}^{(sa)}_{*\to \alpha\beta} &=\{\tau_{i\to j}| s_j=\alpha\ \textrm{and}\ a_j=\beta\},
    \label{eq:tau_sexage2}
\end{align}
where $\alpha\in\{\textrm{M},\textrm{F}\}$ and $\beta\in\{\textrm{0--9},\ldots,\textrm{90+}\}$. As shown in Fig.~\ref{fig:age}(c,d), the effects of age groups on the transmission intervals are consistent with the results considering only the age groups [Fig.~\ref{fig:age}(a)]. Yet the male ``90+'' group tends to have slightly shorter transmission intervals than the female group of the same age for both cases with infectors and infectees.

\subsection{Effects of location and relation on transmission intervals}

To study the effects of location on transmission interval statistics, we define sets of transmission intervals for infectors as follows:
\begin{align}
    \mathcal{T}^{(l)}_{\alpha\to *}=\{\tau_{i\to j}| l_i=\alpha \},
    \label{eq:tau_loc}
\end{align}
where $\alpha\in\{\textrm{Seoul},\ldots,\textrm{Jeju}\}$. In this subsection, we present only the results for infectors as those for infectees are qualitatively the same as those for infectors. As shown in Fig.~\ref{fig:loc}(a), most locations show quite similar statistical properties of $\tau$s. However, Busan, Daegu, and Chungbuk have broader transmission interval distributions than others, while Gangwon and Jeonnam have narrower distributions than others. The similar patterns are observed in Fig.~\ref{fig:loc}(b), where the same transmission interval sets are visualized on the map in terms of the average and standard deviation of $\tau$s. 

We now investigate the combined effects of sex, age group, and location on the transmission interval statistics. For this, we define the sets of transmission intervals for infectors as
\begin{align}
    \mathcal{T}^{(sal)}_{\alpha\beta\gamma \to *}=\{\tau_{i\to j}| s_i=\alpha\ \textrm{and}\
    a_i=\beta\ \textrm{and}\ l_i=\gamma\},
    \label{eq:tau_sexageloc}
\end{align}
where $\alpha\in\{\textrm{M},\textrm{F}\}$, $\beta\in\{\textrm{0--9},\ldots,\textrm{90+}\}$, and $\gamma\in\{\textrm{Seoul},\ldots,\textrm{Jeju}\}$. We generate heatmaps for the averages and standard deviations of transmission intervals in each set $\mathcal{T}^{(sal)}_{\alpha\beta\gamma\to *}$ in Fig.~\ref{fig:loc}(c). We first focus on the age groups from 20s to 60s as they make up a majority of transmission intervals. The broader transmission interval distribution for Chungbuk in Fig.~\ref{fig:loc}(a,b) might be due to relatively high values of $\tau$s of 30--39 year old males and 50--59 year old females in that province. Similar observations can be made for Busan and Daegu, while the opposite tendencies are found for Gangwon and Jeonnam. Sejong is somewhat special in that the city shows a relatively smaller average transmission interval than other locations, yet the transmission interval distribution looks indifferent from many other locations. It may be because the outlying transmission intervals for Sejong are shorter than those in other locations.

\begin{figure}[b!]
  \centering
  \includegraphics[width=\columnwidth]{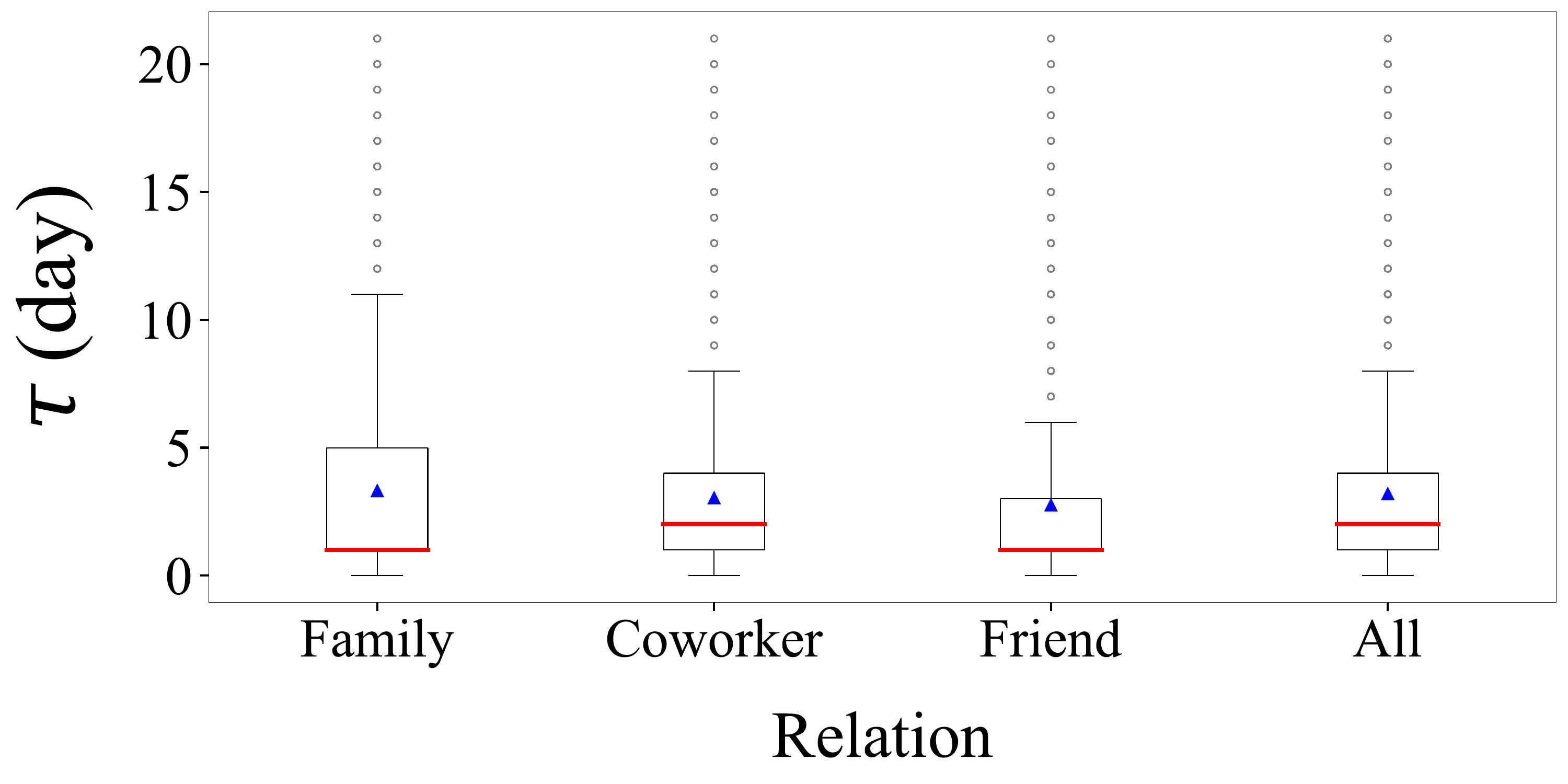}
  \caption{Box-and-whisker plots of $\tau$s for different kinds of relation, i.e., family, coworker, and friend [Eq.~\eqref{eq:tau_rel}], along with the plot for the entire set of transmission intervals, denoted by ``All'', for comparison.
  }
  \label{fig:relation}
\end{figure}

Finally, the nature of relations between infectors and infectees may have an effect on the transmission intervals. We study the following sets:
\begin{align}
    \mathcal{T}^{(r)}_{\alpha} =\{\tau_{i\to j}| r_{i\to j}=\alpha\},
    \label{eq:tau_rel}
\end{align}
where $\alpha\in\{\textrm{Family}, \textrm{Coworker}, \textrm{Friend}\}$. As shown in Fig.~\ref{fig:relation}, it turns out that the transmission intervals between family members are the longest among three relations. This might be related to the previous observation in Fig.~\ref{fig:age}(a,b) that the youngest and oldest age groups tend to have longer transmission intervals than other age groups, possibly due to limited mobility of those groups. In contrast, the transmission intervals between friends are the shortest, which is also consistent with the lowest $\langle \tau\rangle$ and $\sigma_\tau$ between people of similar ages in Fig.~\ref{fig:age}(b). The transmission intervals between coworkers are most similar to those of the entire confirmed cases as the coworkers are expected to have less correlation between their sexes and ages, etc. than family and friend relations.

\subsection{Community transmission intervals}

Communities are one of the most fundamental concepts in network science~\cite{Fortunato2010Community}, and they are known to influence disease spreading among people in a networked population~\cite{Min2013Role, Pastor-Satorras2015Epidemic}. Here we extend the definition of the transmission interval between individuals to the transmission interval at the community level, in particular, using the location information. Precisely, we use basic administrative regions as communities, such as Jongno in Seoul and Seogwipo in Jeju; there are 250 basic administrative regions in Korea.

\begin{figure}[t!]
  \centering
  \includegraphics[width=\columnwidth]{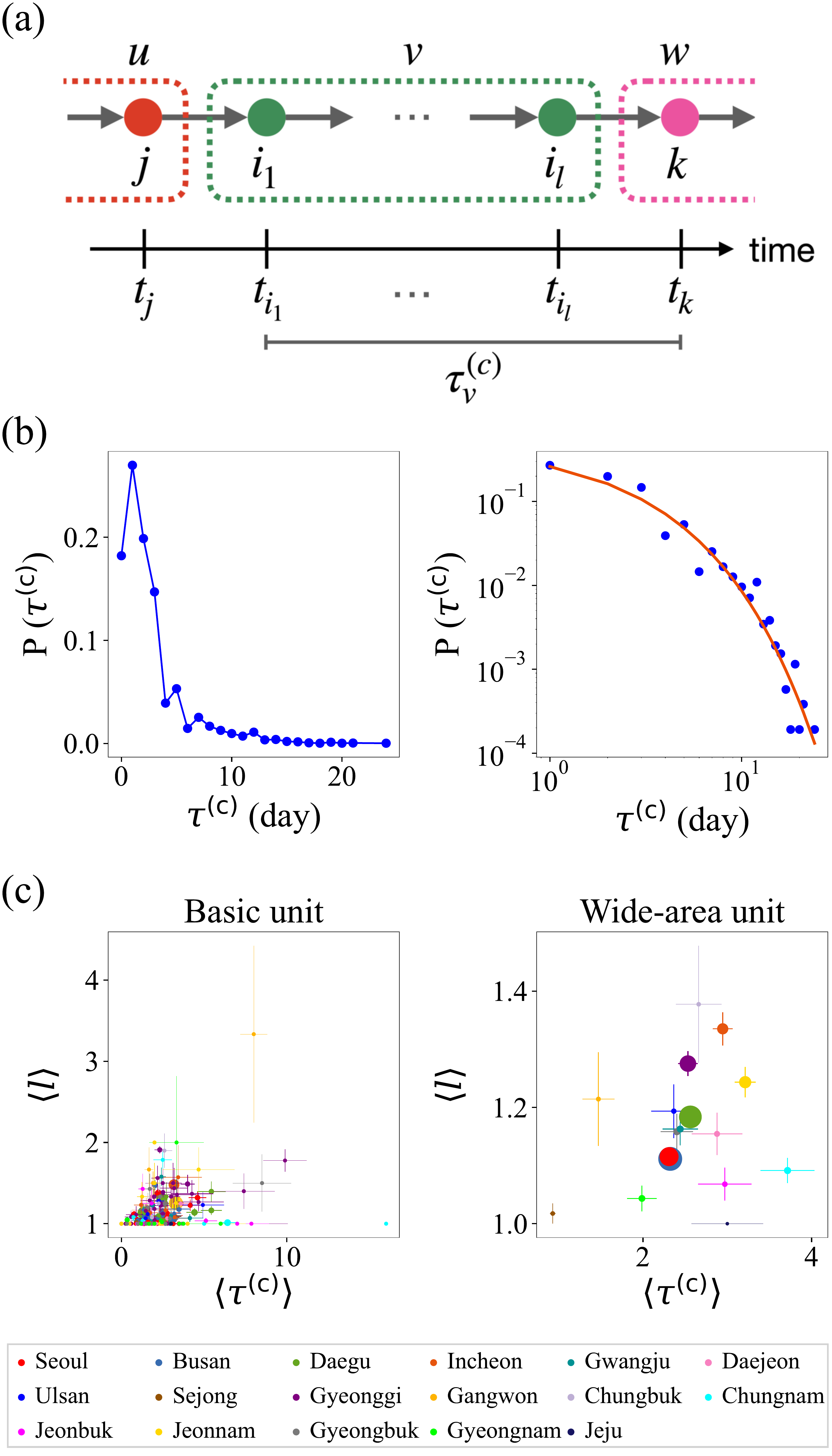}
  \caption{(a) Schematic diagram for the definition of community transmission interval $\tau^{(c)}$, where communities are defined at the level of basic administrative regions (see the main text for details).
  (b) Distribution of community transmission intervals, $P(\tau^{(c)})$, for the entire dataset in a linear plot (left) and in a log-log plot (right), where $\tau^{(c)}$ is measured in days. The orange line in the right panel shows the stretched exponential function fitted to the data.
  (c) Scatter plots of the average transmission interval $\langle \tau^{(c)}\rangle$ versus the average path length of transmission within a community, denoted by $\langle l\rangle$, at both basic (left) and wide-area (right) levels of community. Communities having more $\tau^{(c)}$s are visualized by bigger circles. Error bars denote standard errors in both axes.
  }
  \label{fig:community}
\end{figure}

To define the community transmission interval, we first identify directed paths involving a particular community in the transmission network of individuals. Let us consider three communities, say $u$, $v$, and $w$, and a directed path of transmission from an individual $j$ in $u$ to a sequence of $l$ individuals, i.e., $i_1,\ldots,i_l$, in $v$, finally to another individual $k$ in $w$ [Fig.~\ref{fig:community}(a)], where communities $u$ and $w$ are not necessarily different. Then the community transmission interval is defined as a time interval between the infection times of $i_1$ and $k$:
\begin{align}
    \tau^{(c)}_v\equiv t_k - t_{i_1}=\sum_{m=1}^{l-1 }\tau_{i_m\to i_{m+1}}+\tau_{i_l\to k}.
    \label{eq:tau_c}
\end{align}
That is, the community transmission interval essentially measures how long COVID-19 stays or circulates in the community between arrival and departure. Here $l$ is the number of individuals (or legs) in $v$ in the path, and $l\geq 1$ by definition. One can generically expect to find a positive correlation between $\tau^{(c)}$ and $l$. 

We find that there are 5,218 $\tau^{(c)}$s and their corresponding $l$s available in the dataset. Note that a majority of infection events occur within the same locations~\cite{Kwon2023Clustering}, leaving us with relatively few community transmission intervals. Then we obtain the distribution of community transmission intervals $P(\tau^{(c)})$ as shown in Fig.~\ref{fig:community}(b). The distribution can be fitted with a stretched exponential function, i.e., in a form of $\exp[-(x/x_0)^\mu]$ with $x_0\approx 2(1)$ and $\mu\approx 0.8(2)$. Using Eq.~\eqref{eq:tau_c} statistical properties of $\tau^{(c)}$ can be derived from those of $\tau$ and $l$, e.g., see Ref.~\cite{Jo2013Contextual}.

Then we collect $\tau^{(c)}$s and their corresponding $l$s for each community to calculate averages of them, which are denoted by $\langle \tau^{(c)}\rangle$ and $\langle l\rangle$, respectively. In the left panel of Fig.~\ref{fig:community}(c), we generate a scatter plot of $\langle\tau^{(c)}\rangle$ versus $\langle l\rangle$ for 250 basic administrative regions. As expected, $\langle\tau^{(c)}\rangle$ and $\langle l\rangle$ are overall positively correlated with each other, with the estimated value of Pearson correlation coefficient (PCC) $\approx$$0.26$. In addition, we collect $\tau^{(c)}$s and their corresponding $l$s for communities belonging to the same wide-area administrative regions (or locations in previous subsections). We plot $\langle\tau^{(c)}\rangle$ versus $\langle l\rangle$ for 17 wide-area administrative regions in the right panel of Fig.~\ref{fig:community}(c) to find again the overall positive correlations, with the estimated PCC value $\approx$$0.19$. We also find some outlying cases; e.g., Chungnam shows a larger $\langle\tau^{(c)}\rangle$ but a smaller $\langle l\rangle$ than other regions, implying that COVID-19 circulates within the region longer but over the smaller number of legs than other locations.

\section{Conclusion}

To gain some insights into the effects of demographic factors, such as sex, age, location, and the relation between infectors and infectees, on the disease spreading patterns, we have analyzed the COVID-19 dataset provided by the Korea Disease Control and Prevention
Agency. In particular, we have studied statistical properties of transmission intervals which is defined as a time interval between one person’s infection and their infection to another person. 

We empirically find that transmission intervals are rarely affected by sexes, but they tend to have larger values for the youngest and oldest age groups than other groups. We also find some metropolitan cities or provinces with relatively larger (smaller) transmission intervals than other locations. The nature of relation between infectors and infectees is found to affect the transmission interval statistics such that it takes shorter time for COVID-19 to transmit between friends of similar age than other relations. These empirical findings might help us to better understand dynamical mechanisms of epidemic processes in complex social systems.

\section*{Data availability}
The data used in this study are non-public and have been shared with the COVID-19 mathematical modeling task force team of the Korean Mathematical Society for use in analyzing the basis for establishing the public health policy of the Central Disease Control Headquarters (Korea Disease Control and Prevention Agency).

\begin{acknowledgments}
Y.S. and H.-H.J. acknowledge financial support by the National Research Foundation of Korea (NRF) grant funded by the Korea government (MSIT) (No. 2022R1A2C1007358).
\end{acknowledgments}

%

\end{document}